\documentstyle[pra,aps,latexsym,amssymb,amsmath]{revtex}
\begin{document}
\draft
  \title{Accurate spline solutions of the Dirac equation\\
        with parity-nonconserving potential}
  \author{L.Labzowsky$^{1,2}$ and A.Prozorov$^{1}$}
  \address{$^{1}$  St Petersburg State University,
  198504 Petrodvorets, St Petersburg Russia}
  \address{$^{2}$  Petersburg Nuclear Physics Institute
  188300 Gatchina, St Petersburg Russia }
  \date{\today}
  \maketitle
  \begin{abstract}
     The complete system of the B-spline solutions for the
     Dirac equation with the parity-nonconserving (PNC)
      weak interaction effective
     potential is obtained. This system can be used for the
     accurate evaluation of the radiative
     corrections to the PNC amplitudes
     in the multicharged ions and neutral atoms. The use of
     the scaling procedure allows for the evaluation of the
     PNC matrix elements with relative accuracy $10^{-7}$.
  \end{abstract}

\section{Introduction}

     Consideration of the parity nonconservation (PNC)
     effects in atoms provides an important verification
     of the Standard Electroweak Model in the low energy
     sector. The most important information provides the neutral
     \emph{Cs} atom, because the experimental and the theoretical
     accuracy is the best in this case. The analysis performed
     in~\cite{cW97} and~\cite{sB99} has indicated a deviation
     of the measured weak charge value $Q_{w}$ from that
     predicted by the Standard Model by 2.5 standard
     deviations~$\sigma$. Later this value was diminishing
     and growing up again in a series of
     works~\cite{aD00}-\cite{aV01}. It was understood that the
     radiative corrections to the PNC amplitude  play an
     important role. This question was investigated in
     \cite{wJ01}-\cite{mKvF} where both electron self-energy (SE)
     and vacuum polarization (VP) corrections were evaluated by
     different methods. However, in these calculations $\alpha Z$
     expansion for SE ($\alpha$ is the fine-structure constant,
     $Z$ is the charge of the nucleus), Uehling approximation for
     VP or other approximations were employed. To our mind the
     direct accurate calculations of SE and VP without the use of
     any approximations are desirable. For this purpose we propose
     in this paper the accurate spline solutions of the Dirac
     equation with the weak-interaction PNC potential.

     \par
     Another possible area of application of PNC spline solutions
      could be
     the spectra of the Highly-Charged Ions (HCI). Though no
     experimental results are available here up to now, the
     analysis of the PNC amplitude in the two-electron
     ions can also provide an important test of the
     Standard Model \cite{vG74}-\cite{jS03}.
     The calculation of correlation effects in
     two-electron ions is much easier than in neutral
     atoms, but QED corrections are more important.

     \par
     The rigorous direct way to the calculation of SE and VP corrections
     to the PNC
     amplitude (see~\cite{aP02}) is rather cumbersome. In order to
     avoid the difficulties and receive the accurate results
     we propose to use another way: accurate solutions of the Dirac
     equation with the weak-interaction PNC potential. Having got the
     solutions of the Dirac equation with this potential we can
     obtain self-energy corrections, using the standard methods
     \cite{pM98} - \cite{aM95} based on the spline
     approximation~\cite{wJ88},\cite{cF93}. In this work we solve the
     first part of the problem: the obtaining of the complete basis
     set of accurate solutions of the Dirac equation with the weak
     PNC
     potential. The main difficulty is the extreme smallness of the
     weak interactions in the atomic scale. To overcome this difficulty
     we change the weak
     interaction potential by scaling it in such a way that it becomes
     significant but still small enough to apply the perturbation
     theory. Then we can obtain the necessary PNC matrix elements by
     simple rescaling. Actually our numerical procedure appear to
     be so accurate that the scaling will become necessary only in few
     cases, for the most singular operators and the enhancement of
     the weak interaction potential will not exceed 10 times.

     \section{Dirac equation with PNC weak interaction}

     We employ atomic units $m_{e}=e=\hbar=1$ throughout the
     paper. Consider stationary Dirac equation with the
     weak-interaction PNC potential
     \begin{equation}\label{E:dirac1}
        \left[ c \, \vec{\alpha} \, \vec{p} + V(r) + \beta
        \, c^{2} + V_{w}(r) \right] \, \psi \, (\vec{r})
        = \, E \, \psi \, (\vec{r})
     \end{equation}
     where $\vec{\alpha}$,$\beta$ are Dirac matrices, $\vec{p}$,
     $E$
     are the electron momentum and energy, $c\approx137.03599976$
     is the speed
     of light, $V(r)$ is the Coulomb potential of atomic nucleus
     (pointlike or extended)
     and $V_{w}$ is the weak-interaction PNC potential. This potential
     (see, for example, \cite{iK91}) looks like
     \begin{equation}\label{E:weakpot}
        V_{w} \, = \, \gamma_{5} \, W(r) \, = \, \gamma_{5} \,
        \frac{G_{F}}{2 \, \sqrt{2}} \, Q_{w} \, \rho_{nuc} \, (r)
     \end{equation}
     \begin{equation}\label{E:gam}
        \gamma_{5} \, = \, - \, \left(
        \begin{array}{cc}
           0 & I \\
           I & 0
        \end{array}
        \right) \qquad \qquad
        \begin{array}{l}
           Q_{w} \, = \, - \, N \, + \, Z \, ( \, 1 \, - \, 4 \,
           \sin^{2} \, \theta_{w} \, ) , \\
           \rho_{nuc}(r) \, = \, \rho_{0} \, \left[ \, 1 \, + \,
           \exp \left[ \, \left( r-c \right) / \, a \, \right]
           \, \right]^{-1}
        \end{array}
     \end{equation}
     where $\theta_{w}$ is the Weinberg angle,
     $\sin^{2} \, \theta_{w}\approx0.2230$
     \cite{dG00}, $N$ is the number of neutrons in the
     nucleus, $Z$ is the number of protons, $G_{F}$ is the Fermi
     constant and $\rho_{nuc}(r)$ is the Fermi distribution
     for nuclear charge density. The
     eigenfunctions of the Dirac equation with
     PNC potential have no definite parity, therefore we
     assume
     \begin{equation}\label{E:bispinor}
        \psi \, ( \, \vec{r} \, ) \, = \, \left(
        \begin{array}{c}
           \,  \varphi \, ( \vec{r} \, ) \, \\
           \,  \chi    \, ( \vec{r} \, ) \,
        \end{array} \right)
        \, = \, \frac{1}{r} \left(
        \begin{array}{c}
           \, g^{1}_{njl} \, ( \, r \, ) \, \Omega_{jlM} \, ( \,
           \vec{n} \, ) \, + \, i \, g^{2}_{njl} \, ( \, r \, )
           \, \Omega_{j \bar{l} M} \, ( \, \vec{n} \, ) \, \\
           \, i \, f^{1}_{njl} \, ( \, r \, ) \,
           \Omega_{j \bar{l} M} \, ( \, \vec{n} \, ) \, + \,
           f^{2}_{njl} \, ( \, r \, ) \, \Omega_{jlM} \, ( \,
           \vec{n} \, ) \,
        \end{array} \right)
     \end{equation}
     where $\varphi \, ( \, \vec{r} \, )$ and $\chi \, ( \, \vec{r}
     \, )$ are the upper and lower components of Dirac bispinor,
     $g^{1}_{njl} \, ( \, r \, )$, $g^{2}_{njl} \, ( \, r \, )$,
     $f^{1}_{njl} \, ( \, r \, )$ and $f^{2}_{njl} \, ( \, r \, )$
     are the radial functions, $\Omega_{jlM}$ and
     $\Omega_{j \bar{l} M}$ are spherical spinors with opposite
     parity. The equations for the components $\varphi$, $\chi$ are
     \begin{equation}\label{E:compdir1}
        \, \left( \, E \, - \, V \, ( \, r \, ) \, - \, c^{2} \,
        \right) \, \varphi \, ( \, \vec{r} \, ) \, - \, c \, ( \,
        \vec{\sigma} \, \vec{p} \, ) \, \chi \, ( \, \vec{r}
        \, ) \, - \, W \, ( \, r \, ) \, \chi \, ( \, \vec{r}
        \, ) \, = \, 0
        \end{equation}
     \begin{equation}\label{E:compdir2}
        \, \left( \, E \, - \, V \, ( \, r \, ) \, + \, c^{2} \,
        \right) \, \chi \, ( \, \vec{r} \, ) \, - \, c \, ( \,
        \vec{\sigma} \, \vec{p} \, ) \, \varphi \, ( \, \vec{r}
        \, ) \, - \, W \, ( \, r \, ) \, \varphi \, ( \, \vec{r}
        \, ) \, = \, 0
     \end{equation}
     Inserting~(\ref{E:bispinor})
     in~(\ref{E:compdir1}),(\ref{E:compdir2})
     and separating out the angular variables we get
     \begin{equation}\label{E:sysrad}
        \left[
        \begin{array}{cccc}
           V  (  r  ) & 0 & -  c  \left(
           \frac{d}{dr}  - \frac{\kappa_{j l}}{r}
           \right) & W ( r ) \\[12pt]
           0 & V ( r ) &  W ( r ) &
           c \left( \frac{d}{dr} +
           \frac{ \kappa_{j l}}{r} \right) \\[12pt]
           c \left( \frac{d}{dr} +
           \frac{\kappa_{j \, l}}{r} \, \right) &
           W ( r ) & V ( r ) -  2
           c^{2} & 0 \\[12pt]
           W  (  r  ) & -  c  \left(
           \frac{d}{dr} - \frac{\kappa_{j l}}{r}
           \right) & 0 & V  (  r  )  -  2  c^{2}
        \end{array}
        \right] \, \left[
        \begin{array}{c}
           g^{1} \\ g^{2} \\ f^{1} \\ f^{2}
        \end{array}
        \right] \, = \, \varepsilon \, \left[
        \begin{array}{c}
           g^{1} \\ g^{2} \\ f^{1} \\ f^{2}
        \end{array}
        \right]
     \end{equation}
     where $\varepsilon \, = \, E \, - \, c^{2}$ and
     \begin{equation}\label{E:kappa}
        \kappa_{j \, l} \, = \,
        \begin{cases}
            \quad l  & \text{if     $ \quad j = l
            - \frac{1}{2}$}\\
            - \, ( \, l \, + \, 1 \, )
            & \text{if $ \quad j = l + \frac{1}{2}$}
        \end{cases}
     \end{equation}

     This system of equations for the radial functions was first
     derived in~\cite{aS85}. However in~\cite{aS85} only one certain
     solution of Eq~(\ref{E:sysrad}) was obtained. Our finite goal
     requires the knowledge of the complete system of the solutions
     of Eq~(\ref{E:sysrad}). For this purpose we use the B-spline
     approach~\cite{wJ88},\cite{cF93}.

\section{Parity-nonconserving B-spline Approach}

     Using the Galerkin method (see~\cite{cF84} and~\cite{wJ88}) we
     express the system of equations~(\ref{E:sysrad}) in terms of the
     least action principle.
     We use MIT bag-model boundary conditions~\cite{aC74}, employed
     also in~\cite{wJ88}. In case of PNC functions these conditions
     look like
     \begin{equation}\label{E:boundcon}
     \begin{cases}
     \, g^{1} \, ( \, R \, ) \, = \, f^{1} \, ( \, R \, ) \\
     \, g^{2} \, ( \, R \, ) \, = \, f^{2} \, ( \, R \, )
     \end{cases}
     \qquad
     \begin{cases}
     \, g^{1} \, ( \, 0 \, ) \, = \, 0 \\
     \, g^{2} \, ( \, 0 \, ) \, = \, 0 \; .
     \end{cases}
     \end{equation}
     Here $R$ is the size of a spherical box, where an atom or
     an ion is enclosed.
     We choose action functional $S$ so that condition $\delta S =0$
     leads to Eqs~(\ref{E:sysrad}) and~(\ref{E:boundcon}). Then we present
     the functions $g^{1}, \; g^{2}, \; f^{1} \; \mbox{and} \;
     f^{2}$ in the form of the linear combination of B-splines.  The
     condition $\delta S=0$ reduces to the system of $4n \times 4n$ symmetric
     generalized eigenvalue equations. For constructing the B-spline
     system we use the grid $\quad r_{i}=\rho_{i}^{4}/Z,
     \quad i=0,\ldots,N_{int}$, that was described in~\cite{cF93}.
     In our calculations we use the number of the grid
     intervals $N_{int}=200$, the order of B-splines $k=9$ and the
     parameter, which defines the density of knots near zero
     (see~\cite{cF93}), $h_{set}=0.0167$.

     \par
     In this paper we investigate the hydrogenlike ion $^{238}U^{91+}$.
     The box size was taken to be $R \approx 1.0$ a.u.
     The parameters of the Fermi
     distribution are $c=7.136$ fm and $a=2.3/4\ln3$~\cite{eroh}.
     As a first test of the
     accuracy of our B-spline approach, we set $V(r)=-Z/r$, $N_{w}=0$
      and
     compare the energy values from B-spline approach with
     low-lying energy levels given by the
     Sommerfeld formula. The results are presented in the Table~\ref{T:t1} in a.u.
     Our accuracy of calculation of the energy $10^{-14}$ is the
     limit of accuracy for the calculation with Fortran double
     precision type of variables.

     Another check of the spline accuracy is given in
     Table~\ref{T:t2}, where the matrix elements of the operators
     $r^{m} \; (m=2,1,0,-1,-2)$ are compared on the $2s_{1/2}$ and
     $2p_{1/2}$ spline and exact (point-like nucleus) wave functions without weak
     interaction. The results in Tables~\ref{T:t1}~and~\ref{T:t2} show that
     all the matrix elements $<2s_{1/2} |r^{m}|2s_{1/2} >$
     and \\
     $<2p_{1/2} |r^{m}|2p_{1/2} >$
     apart from the case $m=-2$ can be evaluated with the same
     relative
     accuracy  $10^{-14}-10^{-15}$ as the low-lying energies. Only in case
     of $m=-2$ the accuracy is of order $10^{-7}-10^{-8}$ because of the singularity
     of $r^{-2}$ operator.

     \section{Scaling and the numerical tests}
     As the next step we increase the PNC weak interaction by scaling
     $V_{w}=N_{w} \, \gamma_{5} \, W(r)$, where $N_{w}$ is the
     scaling factor, and investigate the energy spectrum by
     changing $N_{w}$. Here we use the Fermi distribution for
     the nuclear charge density.

     \par
     The scaling problem looks as follows: to find a scaling
     parameter $N_{w}$ for which 1) the contribution of weak interaction
     to the matrix elements of our interest is larger
     than numerical inaccuracy, 2) weak interaction is still
     small enough
     for using the perturbation theory.

     \par
     The Table~\ref{T:t3} presents the PNC weak interaction
     contribution to
     the energy of the states for the different values of $N_{w}$.
     In case $N_{w} \ne 0$ the states do not have a certain parity
     and we use the notations $ n \tilde{s},n \tilde{p}$ formally
     classifying the different states by their origin.
     The
     analysis of the results of calculations shows that
     the dependence $E=E(N_{w})$ is approximately parabolic.
     One can also obtain this
     result from the perturbation theory (PT):
     \begin{equation}\label{E:ptfun2}
       |n \tilde{s} \rangle = |ns \rangle+N_{w} \sum_{mp}
       \frac{\langle
       mp|V_{w}(r)|ns\rangle}{E_{ns}-E_{mp}}|mp \rangle+
       \frac{1}{2} \left( N_{w} \right)^{2} \sum_{m} \sum_{l}
       \frac{ \langle ls|V_{w}|mp \rangle
       \langle mp|V_{w}|ns
       \rangle}{(E_{mp}-E_{ls})(E_{ns}-E_{mp})} |ls \rangle
     \end{equation}
     \begin{equation}\label{E:energy}
       \langle n \tilde{s}| \tilde{H}|n \tilde{s} \rangle-
       E_{ns}= -\left( N_{w} \right)^{2} \, \sum_{m}
       \frac{\left| \langle mp|V_{w}|ns \rangle \right|^{2}}
       {E_{ns}-E_{mp}}
     \end{equation}
     Here the evident equalities $\langle ms|V_{w}|ns \rangle=0$
     and  $\langle mp |H(r)|ns \rangle=0$ were used. The
     right-hand
     side of Eq~(\ref{E:energy}) is also presented in
     Table~\ref{T:t3} for $N_{w}=1$. The comparison shows that we
     can obtain the PNC weak interaction contribution to
     the energy with accuracy $0.03 \; \%$ if we use scaling
     parameter $N_{w}=10^{2} \div 10^{3}$. Here by the numerical
     inaccuracy we undestand the
     B-spline method inaccuracy of calculation of
     nonperturbative energy value presented in the last column of
     Table~\ref{T:t3}. This  leads to the poor accuracy
     of the weak-interaction PNC contributions to the energy,
     obtained with
     the
     low values of scaling parameter $N_{w}=1 \div 10$. For
     $N_{w}>10^{4}$ higher orders of perturbation theory
     become to be not negligible.

     \par
     The second-order PNC weak interaction correction to the
     energy is of no physical interest and was investigated here
     only for the demonstration of the possibilities of the scaling
     procedure. This procedure helps to evaluate the second-order PNC
     weak interaction correction even though this correction is
     beyond the level of accuracy of the spline approximation.

     \par
     We should emphasize that the matrix elements of our interest
     are the matrix elements with linear dependence on weak
     interaction. SE and VP corrections to the PNC amplitude are
     the objects of this type.

     \par
     For testing our system of PNC B-spline functions we calculated
     the set of the matrix elements $\langle 2 \tilde{s}|r^{m}|2
     \tilde{s} \rangle$ with the first non-zero order of
     perturbation expansion, linear in $V_{w}$. The following
     tests were employed:
     \begin{equation}\label{E:rm1}
     \langle 2 \tilde{s}|r^{m}|2 \tilde{p} \rangle =
     \sum_{n} \frac{\langle 2 \tilde{s}|r^{m/2}|n
     \tilde{s} \rangle \langle n \tilde{s}|r^{m/2}|
       2 \tilde{p} \rangle}{E_{2 \tilde{p}}-E_{n
       \tilde{s}}}+\sum_{n} \frac{\langle 2 \tilde{p}|r^{m/2}|n
     \tilde{p} \rangle \langle n \tilde{s}|r^{m/2}|
       2 \tilde{p} \rangle}{E_{2 \tilde{p}}-E_{n
       \tilde{p}}}
     \end{equation}
     \begin{equation}\label{E:rm2}
     \langle 2 \tilde{s}|r^{m}|2 \tilde{p} \rangle
     =\sum_{n} \frac{\langle 2s|r^{m}|ns \rangle \langle ns|V_{w}|
       2p \rangle}{E_{2p}-E_{ns}}+\sum_{n}
       \frac{\langle 2s|V_{w}|np \rangle \langle np|r^{m}|
       2p \rangle}{E_{2s}-E_{np}} \qquad
     \end{equation}
     With these tests we check the completeness of the PNC
     B-spline spectrum  (Eq~(\ref{E:rm1})) and
     the absence of the higher order PNC contributions after the
     introduction of the scaling parameter $N_{w}$ i.e. the
     possibility of rescaling  (Eq~(\ref{E:rm2})). The results are given in
     Table~\ref{T:t4}. The behaviour of the matrix elements with $m=2,1$
     is similar. The most suitable scaling parameters for the calculation
     of these matrix elements are $N_{w}=1$ (the absence of the scaling)
     with relative accuracy
     $1 \cdot 10^{-8}$ and $N_{w}=10$ with relative accuracy
     $5 \cdot 10^{-8}$. The test of completeness of the spectrum
     for these operators gives the relative accuracy
     $1 \cdot 10^{-12}$.  The most
     interesting matrix elements are the ones with
     $m=-1,-2$ because of
     their
     singularity. For $N_{w}=1$ the relative inaccuracy for the calculation
     of both matrix elements is about $1 \cdot 10^{-6}$, for
      $N_{w}=10$ it is $1 \cdot 10^{-7}$. The checking
     of completeness of the basis
     for this operators give the relative accuracy
     $1 \cdot 10^{-7}$ in the worst case. Summarizing these
     tests
     we conclude that the best scaling parameter for
     the calculation of matrix element with linear dependence on weak
     interaction is $N_{w}=10$. The accuracy of this calculation
     is about $1 \cdot 10^{-7}$.

     \par
     Finally, we want to illustrate our method by calculation of
     VP corrections in the Uehling approximation to the matrix element
     of PNC weak interaction on $|2s_{1/2} \rangle $ and $|2p_{1/2}
     \rangle $ functions. Such electroweak radiative
     corrections for the HCI were first obtained
     in~\cite{iB99} by solving Dirac equation with the
     Uehling potential
     \begin{equation}\label{E:uehl}
     V_{u}= \frac{2 \alpha Z}{3 \pi r} \int_{1}^{\infty}
     e^{-2ry/ \alpha } \left( 1+ \frac{1}{2y^{2}} \right)
     \frac{ \sqrt{y^{2}-1}}  {y^{2}} dy \; .
     \end{equation}
     The scheme of calculation of these corrections via PT looks as
     follows
     \begin{align}\label{E:ulpt}
       & \qquad \qquad \langle 2 s'|V_{w}|2 p' \rangle -
       \langle 2 s|V_{w}|2 p \rangle=
       \sum_{n \neq 2} \frac{\langle 2s|V_{u}|ns \rangle \langle ns|V_{w}|
       2p \rangle}{E_{2s}-E_{ns}}+\sum_{n \neq 2}
       \frac{\langle 2s|V_{w}|np \rangle \langle np|V_{u}|
       2p \rangle}{E_{2p}-E_{np}}+    \\
       & + \sum_{n \neq 2} \sum_{l \neq 2}  \left( \frac{\langle
       2s|V_{u}|ns \rangle \langle ns|V_{w}|lp \rangle
       \langle lp|V_{u}|2p \rangle }{(E_{2s}-E_{ns})(E_{2p}-E_{lp})}
        +  \frac{ \langle 2s|V_{u}|ls \rangle \langle ls|V_{u}|ns \rangle
       \langle ns|V_{w}|2p \rangle }{(E_{2s}-E_{ns})(E_{2s}-E_{ls})}
       +  \frac{ \langle 2s|V_{w}|np \rangle \langle np|V_{u}|lp \rangle
       \langle lp|V_{w}|2p \rangle }{(E_{2p}-E_{np})(E_{2p}-E_{lp})}
        \right)  -
         \notag \\
        & \qquad \qquad \qquad - \sum_{n \neq 2}  \frac{\langle 2s|V_{w}|np \rangle
          \langle np|V_{u}| 2p \rangle  \langle 2p|V_{u}| 2p
          \rangle}
          {(E_{2p}-E_{np})^{2}} - \sum_{n \neq 2}
          \frac{\langle 2s|V_{u}|2s \rangle
          \langle 2s|V_{u}|ns \rangle
          \langle ns|V_{w}|
       2p \rangle}{(E_{2s}-E_{ns})^{2}} - \notag \\
       & \qquad \qquad \qquad \qquad \qquad - \sum_{n \neq 2}
       \frac{\langle 2s|V_{w}|
       2p \rangle}{2} \left[ \frac{| \langle 2s|V_{u}|ns \rangle |^{2}}
       {(E_{2s}-E_{ns})^{2}}+\frac{| \langle 2p|V_{u}|np \rangle |^{2}}
       {(E_{2p}-E_{np})^{2}}  \right]  \notag
     \end{align}
     where the notation $|2 s' \rangle$ and $|2 p' \rangle$ means
     that this functions were obtained as a solutions of the Dirac
     equation with the Uehling potential. The right-hand side of
     Eq~(\ref{E:ulpt}) presents PT expansion. We
     include the second order of PT because of its
     significance. Our results for the matrix elements in the
     left-hand side of Eq~(\ref{E:ulpt}) are:
     $\langle 2 s|V_{w}|2 p \rangle=-4.163986 \cdot 10^{-6}$ eV and
     $\langle 2 s'|V_{w}|2 p' \rangle=-4.219566 \cdot 10^{-6}$ eV.
      These
     values are slightly different from the values obtained in~\cite{iB99}
     because of our use of Fermi charge density distribution and
      more modern data for the nuclear radius.
      In total the  left-hand side of Eq~(\ref{E:ulpt}) is equal to
      $-5,5580 \cdot 10^{-8}
      $ eV. The
     calculation in the first order of PT in the right-hand side of
     Eq~(\ref{E:ulpt})  gives
     $-5.5058 \cdot 10^{-8}$ eV.  The evaluation of
     the second order terms yields the value $-5,2277 \cdot
     10^{-10}$ eV and the sum of the first and second orders
     terms is  $-5.5581 \cdot 10^{-8}$ eV. Thus the relative
     discrepancy between the left and right-hand sides of
     Eq~(\ref{E:ulpt}) is about $2 \cdot 10^{-5}$. This is the
     limit of accuracy of the calculation of the difference
     between two matrix elements
     $\langle 2 s|V_{w}|2 p \rangle$ and
     $\langle 2 s'|V_{w}|2 p' \rangle$ that were calculated with
     accuracy $10^{-7}$.

     \par

      For our purposes it  is interesting to solve the
     Dirac equation with the PNC weak
     interaction and to calculate then the matrix element of the
      Uehling potential in a direct way and via PT:
     \begin{equation}\label{E:ulwpt1}
     \langle 2 \tilde{s}|V_{u}|2 \tilde{p} \rangle =
     \sum_{n} \frac{\langle 2s|V_{u}|ns \rangle \langle ns|V_{w}|
       2p \rangle}{E_{2p}-E_{ns}}+\sum_{n}
       \frac{\langle 2s|V_{w}|np \rangle \langle np|V_{u}|
       2p \rangle}{E_{2s}-E_{np}}
     \end{equation}
     The results of comparison of the left- and right-hand sides
     of Eq~(\ref{E:ulwpt1}) are given in Table~\ref{T:t4}. The
     agreement is about $1 \cdot 10^{-7}$ for $N_{w}=1 \div 10$.

     \par
     It may be interesting also to compare the evaluation of the
     Uehling-PNC matrix element in a way given by
     Eq~(\ref{E:ulpt}) and by the application of
     Eq~(\ref{E:ulwpt1}). For this comparison we rearrange the
     terms in the right-hand side of Eq~(\ref{E:ulwpt1}),
     separating out the large contribution, containing small
     denominator $\Delta E^{(2)}=E_{2s}-E_{2p}$. We remind that
     in our calculation we do not include the radiative
     corrections in the Dirac energy values. Therefore $\Delta
     E^{2}
     \neq 0 $ only due to the use of the non-Coulomb field of the
     nucleus (Fermi distribution). Using the smallness of the
     differences $\Delta E^{(n)}=E_{ns}-E_{np}$ we replace
     $E_{ns}$ by $E_{np}$ and vice versa in the right-hand side of
     Eq~(\ref{E:ulwpt1}). Then we arrive at the equality
     \begin{equation}\label{E:ulwpt2}
     \langle 2 \tilde{s}|V_{u}|2 \tilde{p} \rangle -
     \frac{ \left[ \langle 2p|V_{u}|2p \rangle
     -\langle 2s|V_{u}|2s \rangle \right] \langle 2s|V_{w}|
       2p \rangle}{E_{2s}-E_{2p}} \simeq \sum_{n \neq 2}
       \frac{\langle 2s|V_{u}|ns \rangle \langle ns|V_{w}|
       2p \rangle}{E_{2s}-E_{ns}}+\sum_{n \neq 2}
       \frac{\langle 2s|V_{w}|np \rangle \langle np|V_{u}|
       2p \rangle}{E_{2p}-E_{np}} ,
     \end{equation}
     the right-hand side of which coincides with the
     first two terms in the right-hand side of
     Eq~(\ref{E:ulpt}). Evaluation of the left-hand side of
     Eq~(\ref{E:ulwpt2}) gives the value $-5.5057 \cdot 10^{-8}$ eV
     in a good agreement with the first-order PT result in
     Eq~(\ref{E:ulpt}).
     \par
     The second order of PT can be obtained by calculation of the
     following expression
     \begin{align}\label{E:ulwpt3}
      &\qquad \sum_{n \neq 2}
       \frac{\langle 2 \tilde{s}|V_{u}|n \tilde{s} \rangle
       \langle n \tilde{s}|V_{u}|
       2 \tilde{p} \rangle}{E_{2 \tilde{s}}-E_{n \tilde{s}}}+
       \sum_{n \neq 2}
       \frac{\langle 2 \tilde{s}|V_{u}|n \tilde{p} \rangle
       \langle n \tilde{p}|V_{u}|
       2 \tilde{p} \rangle}{E_{2 \tilde{s}}-E_{n \tilde{p}}}
       - \qquad  \\
       &- \frac{ \langle 2s|V_{w}|2p \rangle}{E_{2s}-E_{2p}} \sum_{n \neq 2}
       \left(
       \frac{ \langle 2p|V_{u}|np \rangle \langle np|V_{u}|2p \rangle
        }{E_{2p}-E_{np}}
        -\frac{ \langle 2s|V_{u}|ns \rangle \langle ns|V_{u}|2s \rangle
        }{E_{2s}-E_{ns}} \right)
       \notag \\
       & \qquad - \sum_{n \neq 2}
       \frac{\langle 2s|V_{w}|
       2p \rangle}{2} \left[ \frac{| \langle 2s|V_{u}|ns \rangle |^{2}}
       {(E_{2s}-E_{ns})^{2}}+\frac{| \langle 2p|V_{u}|np \rangle |^{2}}
       {(E_{2p}-E_{np})^{2}}  \right]  \notag
     \end{align}
     For this expression we obtain the value $-5,2239
     \cdot 10^{-10}$ eV.
     Then the sum of the first and second orders
     terms is  $-5.5579 \cdot 10^{-8}$ eV. Then, the
     relative accuracy of calculation of the sum of the first
     and second order terms is the same as in the first method:
     $2 \cdot 10^{-5}$. Here we also lose two orders in accuracy
     due to subtraction of the leading terms in
     Eq~(\ref{E:ulwpt2}).  In any case we can state
     emphasize that the accuracy of calculation of any PNC matrix
     elements is of order $10^{-7}$. Of course, one can
     lose the accuracy of calculation because of subtraction of
     two large values.
     The question of the accuracy of calculation of
     some composite expression should be solved separately in each case.
     As we understand, the calculation of the
     radiative corrections to the emission PNC amplitude should not
     contain any subtractions.
     \par
     Concluding, we can say that the method developed here
     presents a powerful tool for the calculation of any
     corrections to the PNC matrix elements in atoms and ions.
     The recommended scaling parameter is $N_{w}=1$ or $N_{w}=10$
     and the relative accuracy for the calculation of PNC matrix
     elements achieved in all cases is not less
     than $10^{-7}$.
     In particular, this approach can be applied to the calculation of the
     radiative corrections to the emission PNC amplitude with high
     accuracy without any
     approximations ($\alpha Z$ expansion, Uehling potential
     approximation and other). Including in our scheme the
     Dirac-Hartree-Fock potential we can calculate these radiative
     corrections also for neutral atoms, e.g. for $Cs$. It will help
     to solve finally the problem of the possible deviation of
     the measured weak charge value $Q_{w}$ from that
     predicted by the Standard Model.

  \acknowledgements
The authors thank I.Goidenko and O.Yu.Andreev for the help with
the B-spline codes. They are also grateful to G.Soff for drawing
their attention to the paper~\cite{aS85}. The financial support by
the RFBR grant 02-02-16758 and by Minobrazovanje grant E02-3.1-7
is acknowledged.

\begin{table}[p!]
  \centering
  \caption{Comparison of the Sommerfeld and spline spectra in
  absence of $V_{w}$}\label{T:t1}
  \begin{tabular}{ c c c r}
  \qquad State & Energy(B-spl.appr.) & Energy(Sommerfeld) & Relative
  inaccuracy \qquad \qquad
  \\  \tableline
  \qquad $   1s_{1/2}  $ & \: -4861.197895993788 \: & \:
  -4861.197895993730 \: & \:
   1.2E-14 \qquad \qquad \qquad \\ 
  \qquad $2s_{1/2}$ & -1257.395849439810 & -1257.395849439807 &  2.4E-15
  \qquad \qquad \qquad \\ 
  \qquad $2p_{1/2}$ & -1257.395849439792 & -1257.395849439807 & -1.2E-14
  \qquad \qquad \qquad \\ 
  \qquad $2p_{3/2}$ & -1089.611415894742 & -1089.611415894764 & -2.0E-14
  \qquad \qquad \qquad \\ 
  \qquad $3s_{1/2}$ & -539.0933280909740 & -539.0933280909703 &  6.9E-15
  \qquad \qquad \qquad \\ 
  \qquad $3p_{1/2}$ & -539.0933280909630 & -539.0933280909703 & -1.4E-14
  \qquad \qquad  \qquad \\ 
  \qquad $3p_{3/2}$ & -489.0370846743426 & -489.0370846743463 & -7.6E-15
  \qquad \qquad \qquad \\ 
  \qquad $3d_{3/2}$ & -489.0370846743426 & -489.0370846743463 & -7.6E-15
  \qquad \qquad \qquad \\ 
  \qquad $3d_{5/2}$ & -476.2615942332995 & -476.2615942332814 &  3.8E-14
  \qquad \qquad  \qquad \\ 
  \qquad $4s_{1/2}$ & -295.2578381192325 & -295.2578381192252 &  2.5E-14
  \qquad \qquad \qquad \\ 
  \qquad $4p_{1/2}$ & -295.2578381192179 & -295.2578381192252 & -2.5E-14
  \qquad \qquad \qquad \\ 
  \qquad $4p_{3/2}$ & -274.4077572604947 & -274.4077572604874 &  2.7E-14
  \qquad \qquad \qquad \\ 
  \qquad $4d_{3/2}$ & -274.4077572604765 & -274.4077572604874 & -4.0E-14
  \qquad \qquad \qquad \\ 
  \qquad $4d_{5/2}$ & -268.9658771399190 & -268.9658771399118 &  2.7E-14
  \qquad \qquad \qquad \\ 
  \qquad $4f_{5/2}$ & -268.9658771399227 & -268.9658771399118 &  4.1E-14
  \qquad \qquad \qquad \\ 
  \qquad $4f_{7/2}$ & -266.3894469008119 & -266.3894469008155 & -1.4E-14
  \qquad \qquad \qquad \\ 
  \qquad $5s_{1/2}$ & -185.4851885786884 & -185.4851885786193 &  3.7E-13
  \qquad \qquad \qquad \\ 
  \qquad $5p_{1/2}$ & -185.4851885785974 & -185.4851885786193 & -1.2E-13
  \qquad \qquad \qquad \\ 
  \qquad $5p_{3/2}$ & -174.9446126829571 & -174.9446126830371 & -4.6E-13
  \qquad \qquad \qquad \\ 
  \qquad $5d_{3/2}$ & -174.9446126830371 & -174.9446126830371 &  1.0E-15
  \qquad \qquad \qquad \\ 
  \qquad $5d_{5/2}$ & -172.1552518809149 & -172.1552518811695 & -1.5E-12
  \qquad \qquad \qquad \\ 
  \qquad $5f_{5/2}$ & -172.1552518811732 & -172.1552518811695 &  2.2E-14
  \qquad \qquad \qquad \\ 
  \qquad $5f_{7/2}$ & -170.8289368147998 & -170.8289368144324 &  2.2E-12
  \qquad \qquad \qquad \\ 
  \qquad $5g_{7/2}$ & -170.8289368144287 & -170.8289368144324 & -2.2E-14
  \qquad \qquad \qquad \\ 
  \qquad $5g_{9/2}$ & -170.0499341722752 & -170.0499341722061 &  4.1E-13
  \qquad \qquad \qquad
\end{tabular}
\end{table}

\begin{table}
  \centering
  \caption{Comparison of the exact (point-like nucleus) and spline values for
  the matrix elements $<2s_{1/2} | r^{m} | 2s_{1/2}>$,
  $<2p_{1/2} | r^{m} | 2p_{1/2}>$, in absence of $V_{w}$.  }\label{T:t2}
  \begin{tabular}{r c c r}
    \multicolumn{4}{c}{ Matrix element $<2s | r^{m} | 2s>$ in
    a.u.} \\
    \hline

   \qquad m  & Spline functions  & Exact functions
   & Relative inaccuracy \qquad \\  \hline
   \qquad 2 & \: 3.428467651648418E-3 \: & \: 3.428467651648391E-3 \:
    & -6.4E-15 \qquad \qquad
    \\
    \qquad 1 & 5.333841373470161E-2
    & 5.333841373470151E-2 & -2.6E-15
    \qquad \qquad
    \\
    \qquad 0 & 1.00000000000000 &
    1.00000000000000 & 1.0E-16 \qquad \qquad \\
    \qquad -1 & 33.2605595132210
     & 33.2605595132212 & 1.2E-14 \qquad \qquad \\
    \qquad -2 & 8415.21963758472 &
    8415.21963813828 & 6.6E-11 \qquad \qquad \\
    \hline
    \multicolumn{4}{c}{} \\ [-9pt]
    \multicolumn{4}{c}{ Matrix element $<2p \, | r^{m} | 2p>$ in
    a.u.} \\ \hline
    m & Spline functions & Exact functions & Relative inaccuracy \qquad \\  \hline
    2 & 2.276847838830044E-3 & 2.276847838830025E-3 &
    -9.2E-15 \qquad \qquad \\
    1 & 4.246884851731023E-2 & 4.246884851731020E-2 & 7.1E-16
    \qquad \qquad
    \\
    0 & 1.00000000000000 & 1.00000000000000 & 1.0E-16 \qquad \qquad \\
    -1 & 33.2605595132206 & 33.2605595132212 & 1.2E-14 \qquad \qquad \\
    -2 &  2542.94240478497 & 2542.94246617904 & 2.4E-08 \qquad \qquad
  \end{tabular}
\end{table}

\begin{table}
\centering
  \caption{Relative contribution of PNC weak interaction to the energies
  for the different values of $N_{w}$ (the left-hand side of
  Eq~(\ref{E:energy}) divided by nonperturbative energy value) and the relative
  contribution of weak interaction to the
  energies calculated by perturbation theory (the right-hand side
  of Eq~(\ref{E:energy}) divided by nonperturbative energy value).
   The last column shows the relative
  inaccuracy of spline method from Table~\ref{T:t1}.}\label{T:t3}
  \begin{tabular}{r r r r r r r r}
      $  N_{w} $ & 1 \qquad \qquad & 10 \qquad \qquad
    & $10^{2}$ \qquad \qquad &
     $10^{3}$ \qquad \qquad & $10^{4}$ \qquad \qquad
    & PT value \qquad & $\delta_{spl}$ \qquad \\  \hline
    $ \ 1\tilde{s}_{1/2}  $ & -2.65322E-13 & -2.94537E-11
      & -2.95031E-09  & -2.95046E-07 & -2.96521E-05
          & -2.95031E-13   & 2.4e-15  \\
    $2\tilde{p}_{1/2}$ & -7.13101E-13 & -3.13915E-11 & -3.14292E-09
    & -3.14285E-07  &   -3.13759E-05 & -3.14291E-13
          &    -1.2E-14   \\
    $2\tilde{s}_{1/2}$ & 6.36938E-14 &  6.64008E-12
    & 6.57424E-10 &  6.57067E-08 &   6.39177E-06 & 6.57255E-14  & -2.0E-14
    \\
  \end{tabular}
\end{table}

\begin{table}
\centering
  \caption{The matrix elements $<2 \tilde{s}_{1/2} | r^{m} | 2 \tilde{p}_{1/2}
  >$ ($m$=2,1,-1,-2) calculated by different methods with the different
  values of the scaling parameter $N_{w}=10^{r}$. The first line
  in the every box for a certain $m$ and $r$ values corresponds to
  the direct evaluation of the matrix element with B-spline
  solution of the Dirac equation with PNC weak interaction. The second
  line corresponds to the use of the formula Eq~(\ref{E:rm1}) in the text.
  The PT line corresponds to the perturbation theory evaluation via
  the formula Eq~(\ref{E:rm2}) with $N_{w}=1$.
  (in a.u.) The last column presents the results of evaluation
  of the Uehling matrix element with the same methods. }\label{T:t4}
  \begin{tabular}{  c  |  c  |  c  |  c  |  c  |  c  }
     r  & \multicolumn{4}{c|}{m} & $\langle 2
     \tilde{s} |V_{u}| 2 \tilde{p} \rangle$ eV
      \\ \cline{2-5}
        & 2 & 1 &   -1 & -2 &
         \\ \hline
     0  & \: -1.971129293262E-8 \: & \: -1.858655956533E-7
     \:   & \: -1.914713446350E-7
     \: & \: -8.871229471602E-2 \: &
     1.75928086582129E-6 \\ \cline{2-6}
        & -1.971129293278E-8  & -1.858655956541E-7
           &  -1.914713461356E-7
            &  -8.871223340185E-2 &
            1.75928086582128E-6 \\ \hline
     1  & -1.971129190577E-7  & -1.858655859887E-6
       & -1.914710897826E-6
        & -8.871235842834E-1  &
        1.75928077138616E-5 \\ \cline{2-6}
        & -1.971129190573E-7  & -1.858655859884E-6
            & -1.914710899371E-6
             & -8.871236585343E-1 &
             1.75928077138618E-5  \\ \hline
     2  & -1.971122272187E-6  & -1.858649339848E-5
       & -1.914707122846E-5   &
     -8.87120565454E+0  &
     1.75927469110662E-4  \\ \cline{2-6}
        & -1.971122272187E-6  & -1.858649339848E-5
           & -1.914707122268E-5
          & -8.87120593112E+0  &
          1.75927469110664E-4 \\ \hline
     3  & -1.970430890847E-5  & -1.857997762638E-4
         & -1.914283967935E-4
          & -8.86823300844E+1 &
           1.75866695639621E-3  \\ \cline{2-6}
        & -1.970430890846E-5  & -1.857997762637E-4
            & -1.914283967449E-4
             & -8.86823302428E+1 &
             1.75866695639620E-3   \\ \hline
     4  & -1.904352098651E-4  & -1.795723793234E-3
        & -1.874284426839E-3
          &  -8.58437274571E+2 &
          1.70059752407138E-2 \\ \cline{2-6}
        & -1.904352098651E-4  & -1.795723793234E-3
            & -1.874284426869E-3
             & -8.58437274521E+2 &
             1.70059752407139E-2    \\ \hline
            & & & & & \\ [-5pt]
     PT & -1.971129278718E-8  & -1.858655943008E-7
         & -1.914711189050E-7
          & -8.871235465273E-2 & 1.75928083527704E-6 \\ [4pt]
  \end{tabular}
\end{table}

\end{document}